\documentclass[aps,twocolumn,prb,groupedaddress]{revtex4-1}
\usepackage[version=3]{mhchem}
\usepackage{graphicx}
\usepackage{dcolumn}
\usepackage{epstopdf}
\usepackage{gensymb}
\newcolumntype{d}[1]{D{.}{.}{#1}}
\newcommand{\SCWO}{\ce{Sr2CuWO6}}
\newcommand{\Neel}{N\'{e}el}
\newcommand{\AFReflection}{$\left({}\frac{1}{2},0,\frac{1}{2}\right){}$}
\newcommand{\ts}{\textsuperscript}
\usepackage{verbatim}

\begin{document}

%\preprint{}

\title{From 3D static order to 2D dynamic correlations in the tetragonal double perovskite \ce{Sr2CuWO6}}

\author{Oliver J. Burrows}
\affiliation{CSEC and School of Chemistry, University of Edinburgh, UK}

\author{G\o{}ran J. Nilsen}
\affiliation{Institut Laue-Langevin, 72 Avenue des Martyrs, Grenoble, France}
\affiliation{ISIS, Rutherford Appleton Laboratory, Harwell Oxford, UK}

\author{Emmanuelle Suard}
\affiliation{Institut Laue-Langevin, 72 Avenue des Martyrs, Grenoble, France}

\author{Mark Telling}
\affiliation{ISIS, Rutherford Appleton Laboratory, Harwell Oxford, UK}

\author{J. Ross Stewart}
\affiliation{ISIS, Rutherford Appleton Laboratory, Harwell Oxford, UK}

\author{Mark A. de Vries}
\email[]{m.a.devries@ed.ac.uk}
\affiliation{CSEC and School of Chemistry, University of Edinburgh, UK}

\date{\today}

\begin{abstract}
\SCWO is a tetragonal ($I4/m$) double perovskite with diamagnetic
W$^{6+}$ (5d$^0$) and (Jahn-Teller-active) Cu$^{2+}$ (3$d^9$) in a
rocksalt-ordered arrangement on the B sites, modelling an anisotropic
fcc-lattice antiferromagnet with $S=1/2$. We have studied the magnetic
structure, dynamics and thermodynamic properties of this material
above and below the \Neel{}-ordering transition at $T_{\textrm{N}} =
24$~K. Muon spin relaxation spectroscopy, neutron diffraction and
neutron spectroscopy experiments show that the transition at
$T_{\textrm{N}}$ on heating is from static 3D type-2
antiferromagnetism to 2D dynamic correlations that remain detectable
up to 100~K. Above $T_{\textrm{N}}$ the muon relaxation is described
by a compressed exponential decay, revealing a strong correlation
between the spin fluctuation rate and the 2D correlation length. The
low-temperature muon data and the entropy release around
$T_{\textrm{N}}$ as obtained from the heat capacity data point to
small magnetic domains, probably of the order  of 60~\AA{}. We note
that the material behaves more like a quasi-2D square lattice
antiferromagnet than could be expected considering its structure
only. A possible explanation might lie in the unusual strong
spin-orbit-coupled magnetism of the W 5$d$ levels, as the dominant
exchange interaction is via the Cu-O-W-O-Cu pathway. 
\end{abstract}

%\keywords{}

\maketitle

\section{Introduction}
Double perovskite Mott-Hubbard insulators have attracted much
attention because of their interesting, often geometrically
frustrated, magnetic topologies as well as the orbital degrees of
freedom associated with high symmetry of the magnetic metal-ion
environment~\cite{chen_2010_exotic}. These features are exemplified by
the $B$-site rocksalt-ordered double perovskites in which one of the
$B$-sites contain orbitally-degenerate $4d$ or $5d$ transition metal
ions with strong spin-orbit coupling ( Mo$^{5+}$, Ru$^{6+}$,
Re$^{6+}$, Os$^{7+}$ and W$^{5+}$), which can be modelled as fcc
lattices of antiferromagnetically-coupled spin-orbital moments with
near-neighbour and next-nearest neighbour interactions~\cite{witczak-krempa_2013_correlated}. Against this backdrop
the double perovskites \ce{Ba2CuWO6} and
\ce{Sr2CuWO6}~\cite{blasse_1965_new} provide model systems containing (
perhaps by now reassuringly familiar) antiferromagnetically-coupled
Cu$^{2+}$ $J=S=1/2$ spin moments, the W cations being in the nominally
non-magnetic 6+ state~\cite{blasse_1965_new, gateshki_2003_x-ray,
  vasala_2012_synthesis-a}. The Ba-analogue (\ce{Ba2CuWO6}) has previously been shown to
have a long-range ordered type-2 \Neel ground state with $T_N =
28$~K~\cite{todate_2007_magnetic}, but because of the absence of any
anomaly or field-cooled zero-field cooled splittings in the magnetic
susceptibility \ce{Sr2CuWO6} was put forward as a  potential quantum
spin liquid~\cite{vasala_2012_synthesis-a}. 

\begin{figure}
\includegraphics[width=8.6cm]{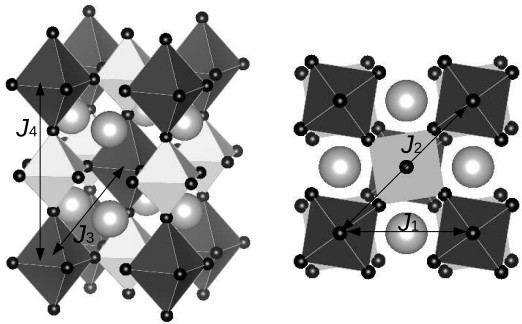}
\caption{The \SCWO structure as seen from the side ($c$-axis pointing
  up) and from the top (along the $c$-axis). The arrows labelled
  $J_1 \ldots J_4$ are the exchange pathways as described in the
  text. $J_1$ coincides with the $a$-axis ($=~b$-axis) and $J_4$ coincides
  with the $c$-axis. 
\label{fig:SCWO_structure}}
\end{figure}

\ce{Sr2CuWO6} undergoes a Jahn-Teller distortion from cubic
($Fm\overline{3}m$) to tetragonal ($I4/mmm$) at
920$^{\circ}$C, leading to an elongation of the $c$-axis ($c/a\sqrt{2}
= 1.1$ for the body-centered tetragonal
cell)~\cite{gateshki_2003_x-ray}. At 670$^{\circ}$C there is a further
transition from $I4/mmm$ to $I4/m$ with a modest rotation of
the octahedra around the $c$-axis, leading to the structure shown in
Fig.~\ref{fig:SCWO_structure}. Subsequent studies on the magnetic
properties of \ce{Sr2CuWO6} have focussed on the $I4/m$
phase. The magnetic Cu$^{2+}$ $d_{x^2-y^2}$ orbital is aligned in the
$a-b$ plane, causing some anisotropy between the in-plane and
out-of-plane exchange constants. Figure~\ref{fig:SCWO_structure} shows
the four most important exchange interaction pathways; $J_1$ and $J_3$
are via Cu-O-O-Cu (90$^{\circ}$) bonds in-plane and out-of plane
respectively. $J_2$ and $J_4$ are via the Cu-O-W-O-Cu (180$^{\circ}$)
pathways in the $a-b$ plane and along the $c$-axis, respectively. As
remarked earlier~\cite{vasala_2012_synthesis-a}, there is potential
for geometrical frustration due to competing interactions along the
different pathways. A recent $\mu$SR experiment has shown that there
is long-range order below $T_N =
24$~K~\cite{vasala_2014_characterization}, roughly an order of
magnitude below the Weiss temperature. In the same study the exchange
constants are estimated using DFT calculations combined with O K-edge
X-ray absorption spectroscopy, yielding (strongest first): $J_2 =
-7.47$~meV, $J_4 = -4.21$~meV, $J_1 = -1.2$~meV and $J_3 =
-0.03$~meV~\cite{vasala_2014_characterization}. While this is at first
sight consistent with the crystallographic structure, the implied 3D
antiferromagnetism is difficult to reconcile with the magnetic
properties which have previously been suggested to point to quasi-2D
magnetism~\cite{vasala_2014_characterization}.

The Mermin-Wagner theorem~\cite{mermin_1966_absence} forbids magnetic
ordering for $T>0$ for any layered system where the interactions
between $S=1/2$ spins are strictly in the plane and short-ranged. In
the case of the 2D square  lattice of $S=1/2$ spins with
antiferromagnetic near-neighbour interactions the $T=0$ ground state
is expected to be long-range (N\'{e}el) ordered~\cite{Manousakis:91}
and in all known cases a 3D N\'{e}el-ordered state sets in at some
finite temperature due to 3D interlayer interactions, which are often
orders of magnitude smaller than the dominant in-plane exchange
interactions~\cite{keimer_1992_magnetic}. Reduced ordered moments due
to quantum fluctuations have been observed in physical realisations
including copper formate tetradeuterate
(CFTD)~\cite{ronnow_2001_spin}. The spin-wave dispersion exhibits
distinctive non-classical features indicative of emergent $S=1/2$
quasiparticle
excitations~\cite{dalla_piazza_2015_fractional}. Furthermore, in many
of these systems, including \ce{La2CuO4}~\cite{keimer_1992_magnetic},
\ce{Sr2CuO2Cl2}~\cite{greven_1995_neutron} and
CFTD~\cite{ronnow_1999_high}, strong dynamic correlations have been
found to persist far above the ordering temperature.  Long-ranged 2D
dynamic correlations in the $a-b$ planes, above $T_{\textrm{N}}$, have been
  hypothesised to explain the absence of any clear signatures of
\Neel{} ordering in the bulk magnetism in
\ce{Sr2CuWO6}~\cite{vasala_2014_characterization}.    

%Consideration of the theoretically predicted exchange constants, as
%provided by Vasala~\emph{et al.}~\cite{vasala_2014_characterization}
%using \emph{ab initio} calculations combined with X-ray absorption
%spectroscopy on the oxygen K-edge, complicates this picture.
%According to this study the dominant exchange interactions are
%along the Cu-O-W-O-Cu pathways $J_2$ and $J_4$, with $-7.47$~meV and $-4.21$~meV in the
%$a-b$ plane and along the $c$-axis, respectively. The exchange
%interactions along the 90$^{\circ}$ Cu-O-O-Cu pathways $J_1$ and $J_3$
%are significantly weaker, with $-1.2$~meV (in-plane) and $-0.03$~meV
%(out-of-plane). If these exchange interactions are correct then the
%system cannot be described as quasi-2D and there is then no
%justification for the proposed 2D dynamic correlations developing well
%above $T_{\textrm{N}}$. 

Notwithstanding all the work being carried out on \SCWO, the magnetic
ground state order is not yet known, nor is there any other than
circumstantial evidence of dynamic correlations above the \Neel
temperature. This paper addresses both aspects with the aid of neutron
powder diffraction, neutron spectroscopy, muon spectroscopy and heat
capacity measurements. The experimental results confirm earlier
suggestions that this material behaves like a quasi-2D
antiferromagnet. An attempt is made to reconcile this behaviour with
the apparently solidly 3D magnetic topology of this model system
without invoking the concepts of geometrical frustration, which we
argue do not play a role in \SCWO.

\section{Experimental Details}
\SCWO{} was synthesised using standard solid state
techniques. Stoichiometric quantities of \ce{SrCO3}, \ce{CuO} and
\ce{WO3} were ground in an agate mortar in air, before being pressed
into a pellet and calcined at $900$~\degree{}C for two hours in
air. The resulting compound was re-pelleted, then sintered at
$1200$~\degree{}C (again in air) for $12$~hours. The product was
characterised using powder X-ray diffraction, and the sintering step
was repeated until phase purity was achieved. 

DC magnetic susceptibility measurements took place on a Quantum Design
MPMS between $2$~K and $400$~K. Magnetization was measured as a
function of temperature at a range of fields up to $7$~T. 

Heat capacity was measured on a Quantum Design PPMS, at temperatures
between 2 and 300~K with 0.2~K steps between 20 and 30~K. A thin
square pellet ($1.5$~mm by $1.5$~mm, mass  $9.5$~mg) of sintered \SCWO
was affixed using Apiezon N grease (Dow Corning). The heat capacity of
the grease was measured prior to the heat capacity of sample + grease,
so that the heat capacity of the small sample could be extracted.

High-resolution neutron diffraction was carried out in the D2B
beamline at the ILL on a $2$~g sample in a cylindrical vanadium
can. At
$3.5$~K the neutron diffraction data collected at wavelengths
$1.594$~\AA{} (germanium monochromator $\left({}335\right){}$
reflection) and $2.398$~\AA{} ($\left({}331\right){}$ reflection) were
refined together. At $50$~K a wavelength of $2.398$~\AA{} was used,
and at $298$~K $1.594$~\AA{}. The uncollimated neutron beam was used
with maximum slit openings.

The X-ray and neutron diffraction data were refined using the
\textsc{GSAS/ExpGUI}\cite{larson_1994_general,toby_2001_expgui}
software. Magnetic analysis of neutron diffraction was carried out
using \textsc{Fullprof}~\cite{rodriguez-carvajal_1993_recent} and the
Bilbao Crystallographic Server~\cite{}.

Muon spin relaxation spectroscopy was carried out on the MuSR beamline at
ISIS. Powdered sample was placed in a disc-shaped holder, and measured
across a range of temperatures from  $1.92$~K to $199$~K and in
longitudinal fields from $0$~G to $200$~G. Measurements for data
presented here were taken for $38.4$ million
positrons~\cite{positrons} detected in each 
case. The polarization $P_z$ is calculated from the
asymmetry $A(t)$ using
\begin{equation}
P_z(t) = A(t) / A(t=0)
\end{equation}
where $A(t) = B(t) - \alpha F(t)$ and $\alpha$ is a factor correcting for
slight differences in efficiency of detecting forwards ($F(t)$) and
backwards ($B(t)$) emitted positions. Above 27~K $A(t=0)$ was equated
to the first data bin. Below 35~K it was clear from the sharp
reduction of $A(t=0)$ compared to the higher temperature data that
significant muon relaxation took place even within the ``$t=0$'' bin
and hence for $T<35$~K a constant $A_T(t=0) = A_{\textrm{35~K}}(t=0)$ was used.

Finally, inelastic neutron spectroscopy measurements were done on the MARI
beamline at ISIS on a $60$~g sample. The sample was placed in an 
aluminium foil envelope, which was rolled into annular configuration 
and placed in a 50~mm aluminium can. The
incident energy was $30$~meV, and measurements were taken at
temperatures between $5$~K and $200$~K, using the thin-tail CCR. 

\section{Results}
\subsection{Bulk measurements}
\begin{figure}
\includegraphics[width=8.6cm]{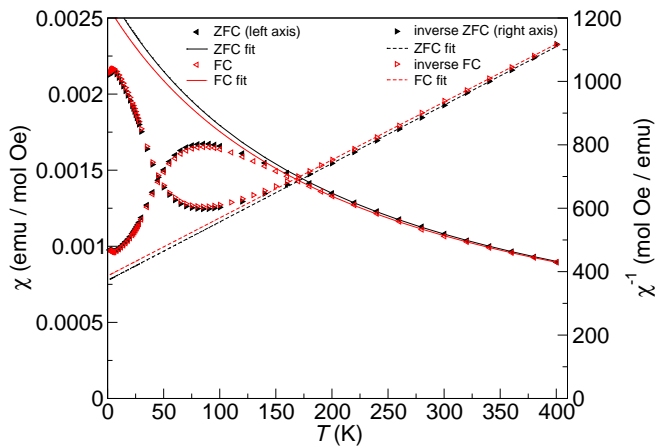}
\caption{Magnetic susceptibility (left axis, open symbols) and inverse
  susceptibility (right axis, filled symbols) of \SCWO{} in
  zero-field-cooled (circles) and field-cooled (square) measurements,
  with fits to Curie-Weiss paramagnetism above
  $170$~K.\label{fig:SCWOSusc}} 
\end{figure}

Magnetometry measurements indicate that the high-temperature state,
above 100~K, is Curie-Weiss paramagnetic. The Curie constant $C = 0.54
\pm{}0.01$~emu~mol\textsuperscript{-1}~K\textsuperscript{-1}
corresponds to an effective moment $\mu_{\textrm{eff}} =
2.08~\mu_{\textrm{B}}$. The effective moment for Cu$^{2+}$ is usually
around $1.90~\mu_{\textrm{B}}$ and we do not have an explanation ready
for the large effective moment observed here, except that perhaps
there is some (minor) hybridisation between the Cu and W cations. This
could cause an increased moment on the Cu (Cu$^{3+}$ mixing in) as
well as a small moment on the W, with W$^{5+}$ mixing in.  The Weiss
temperature $\theta_{\textrm{CW}}$ is $-210\pm{}10$~K
(Fig. \ref{fig:SCWOSusc}). This differs from the Weiss constants of
$-116$~K and $-118$~K previously reported by Vasala \emph{et
  al.}~\cite{vasala_2012_synthesis} and Iwanaga~\emph{et
  al.}~\cite{iwanaga_1999_crystal}, and is closer to the value of
$-230$~K reported by Blasse~\cite{blasse_1965_new}. There is a broad
maximum just below 100~K that has been associated with the build-up
of short-ranged correlations~\cite{vasala_2012_synthesis} and then
again a slight upturn below 10~K that could be  due to paramagnetic
impurities. No anomalies and no splitting between field-cooled and
zero-field-cooled data are observed. This has previously led to
speculation that \ce{Sr2CuWO6} has a 2D spin-liquid magnetic ground
state~\cite{vasala_2012_synthesis}. 

\begin{figure}
\includegraphics[width=8.6cm]{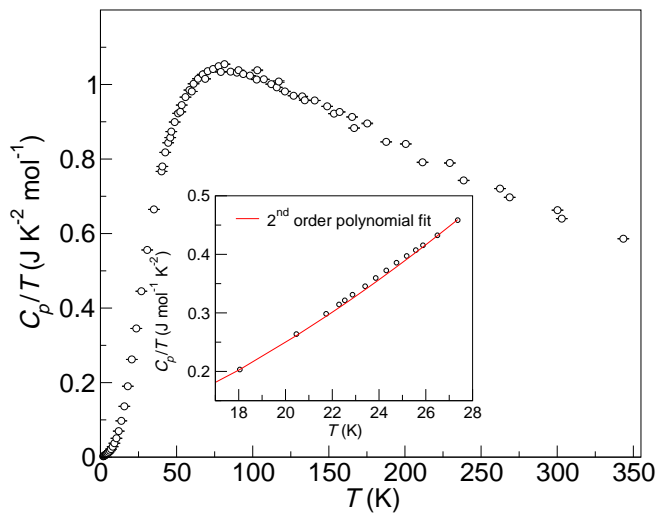}
\caption{Heat capacity of \SCWO{}. Inset shows the region where a weak feature
  caused by the ordering seen in muSR is expected, compared to a
  background derived by fitting a 2\ts{nd}-order polynomial to the
  $C_p/T$ data surrounding the ``peak''.\label{fig:SCWOHeatCap}} 
\end{figure}

\begin{figure}
\includegraphics[width=8.6cm]{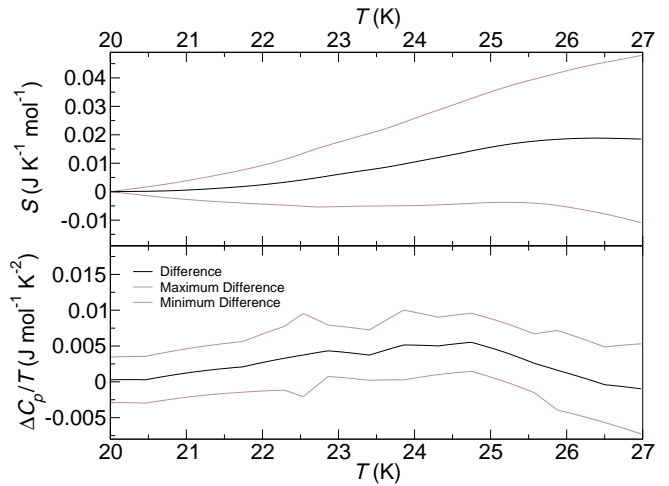}
\caption{Estimate of the magnetic entropy release at the transition
  (top) as a function of temperature (top) based on the difference
  with the fitted background heat capacity (bottom). The minimum and
  maximum difference give help to give an indication of the accuracy
  in the estimated entropy release. It is clear from this data that we
  can only provide an upper bound to the entropy release at the
  magnetic ordering transition. 
\label{fig:Entropy}}
\end{figure}

Our initial heat capacity data did not reveal any magnetic anomalies
in the heat capacity (Fig.~\ref{fig:SCWOHeatCap}). Following the muon
spin relaxation spectroscopy experiment by Vasala \emph{et al.}
revealing a transition at 24~K, more detailed measurements were done
between 21 and 28~K. These data revealed a very weak feature
(Fig.~\ref{fig:SCWOHeatCap}(inset) and Fig.~\ref{fig:Entropy}) and the
entropy release at the transition (on heating through it from 20 to
26 K) is estimated at $0.3\pm0.2$\% of the total magnetic entropy of
$R~\mathrm{ln}(2)= 5.76$~J~K$^{-1}$~mol$^{-1}$ (Cu$^{2+}$)$^{-1}$ 
(Figure~\ref{fig:Entropy}). 
%0.035/8.314*log(2)*100=0.3%

\subsection{Structural probes}
\begin{figure*}
\includegraphics[width=14cm]{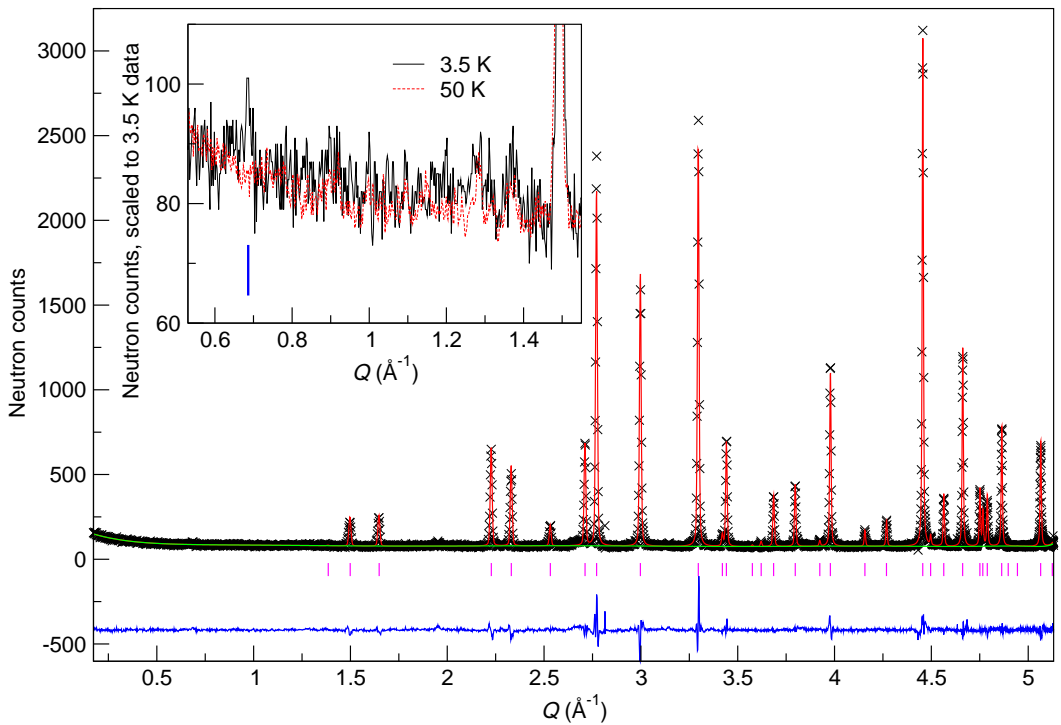}
\caption{Neutron diffraction pattern of \SCWO{} at $3.5$~K, neutron
  wavelength $2.398$~\AA{}. Inset: Comparison of $3.5$~K (solid line)
  and $50$~K (dashed line) neutron diffraction patterns, showing the
  magnetic peak at $Q=0.69$~\AA$^{-1}$. The blue vertical line
  indicates the position of the $\left(
  \frac{1}{2},0,\frac{1}{2}\right)$ magnetic reflection. \label{fig:SCWONDBT}} 
\end{figure*}

The room-temperature neutron diffraction pattern of \SCWO{} is shown
in Fig.~\ref{fig:SCWONDBT}. The pattern was refined in the space group
$I4/m$, as described in previous studies
\cite{todate_2007_magnetic,
  vasala_2012_synthesis}. Table~\ref{tab:SCWORefinements} gives all
details of the present refinement which is in good agreement with
previous data.  The data at $3.5$~K show a very small additional peak
at $Q=0.68$~\AA$^{-1}$ (Fig. 5(inset)). Comparison with a measurement
at 50~K, as shown in the inset of Fig.~\ref{fig:SCWONDBT}, confirms
that the reflection appearing at $0.68$~\AA$^{-1}$ is the only
magnetic Bragg peak within the error bars of our measurements. The
temperature dependence of the anisotropic displacement constants $B$
for Cu and W is anomalous though the 50~K data should not be
considered here as that data is with poorer statistics. That only the
$B$ of the Cu and W cations increases at low temperatures points to a
magnetic effect, perhaps related to the small magnetic domains in the
ordered state. 

\begin{table}%[H] add [H] placement to break table across pages
\caption{Refinement results of \SCWO{} at different
  temperatures. $3.5~K$ and $50~K$ data were refined from neutron
  diffraction data with neutrons of wavelength $2.398$~\AA{}; room
  temperature data were refined from neutron diffraction data with
  wavelength $1.594$~\AA{}. \label{tab:SCWORefinements}} 
\begin{ruledtabular}
\begin{tabular}{ld{3.8}d{3.8}d{3.8}}
\textit{T} (K)                                       & \multicolumn{1}{c}{3.5}           & \multicolumn{1}{c}{50}           & \multicolumn{1}{c}{300} \\
\textit{a} (\AA{})                          & 5.40965(4)    & 5.41065(3)   & 5.42902(3)    \\
\textit{c} (\AA{})                          & 8.41368(7)    & 8.41397(6)   & 8.41614(8)    \\
\textit{V} (\AA{}\textsuperscript{3})                    & 246.221(5)    & 246.320(4)   & 248.059(4)    \\
Sr (0, 0.5, 0.25)                                    &     &    &     \\
\textit{B} (\AA{}\textsuperscript{2})                  & 0.11(4)    &  0.30(4)  &  0.44(2)   \\
Cu                                                   &     &    &     \\
\textit{B} (\AA{}\textsuperscript{2})                  & 0.33(7)   & 0.64(7)   & 0.13(4)    \\
W                                                    &     &    &     \\
\textit{B} (\AA{}\textsuperscript{2})                  & 0.42(9)    & 0.45(10)   &  0.07(6)   \\
O1 (\textit{x}, \textit{y}, 0)                                                  &     &    &     \\
\textit{x}                                           & 0.2892(4)    & 0.2893(4)   & 0.2869(3)    \\
\textit{y}                                           & 0.2024(4)    & 0.2013(4)   & 0.2073(3)    \\
\textit{B} (\AA{}\textsuperscript{2})                  & 0.21(3)     & 0.32(3)   &  0.38(2)   \\
O2 (0, 0, \textit{z})                                                   &     &    &     \\
\textit{z}                                           & 0.2277(2)    & 0.2272(2)   & 0.2269(2)    \\
\textit{B} (\AA{}\textsuperscript{2})                  & 0.02(5)    &  0.15(6)   &  0.73(3)   \\ 
$R_p$ (\%{})                                 & 7.09         & 4.24        &  4.97   \\
$R_{wp}$ (\%{})                                &  8.96        & 5.75        &  6.80   \\
$R_{exp}$ (\%{})                                &  5.53        & 1.64        & 2.50    \\
$\chi{}^2$                                           & 4.785        & 12.39       & 7.498    \\
\end{tabular}
\end{ruledtabular}
\end{table}

For comparison, the neutron diffraction pattern of \ce{Ba2CuWO6} in
the magnetically-ordered state reveals two
peaks~\cite{todate_2007_magnetic}, the first like in \ce{Sr2CuWO6} at
0.68~\AA$^{-1}$, plus an additional reflection at 1.24~\AA$^{-1}$. The
magnetic reflections in \ce{Ba2CuWO6} were indexed as
$\left({}\frac{1}{2},\frac{1}{2},\frac{1}{2}\right){}$ and
$\left({}\frac{1}{2},\frac{1}{2},\frac{3}{2}\right){}$ on a
face-centred unit cell, corresponding to
$\left({}\frac{1}{2},0,\frac{1}{2}\right){}$ and
$\left({}\frac{1}{2},0,\frac{3}{2}\right){}$ in the body-centred
tetragonal setting, respectively.  

\begin{figure}
\includegraphics[width=6.5cm]{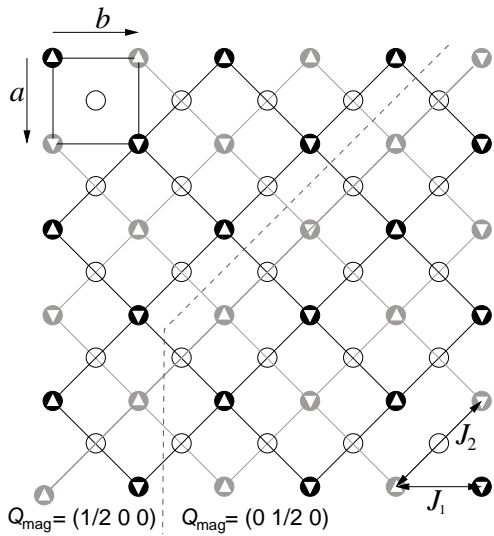}
\caption{A single layer with W$^{6+}$ (open circles) and Cu$^{2+}$ (circles
  with up/down triangle, i.e. spin inside) showing the spin
  configuration in a [001] $a-b$ plane. The same structure is also
  repeated in the [110] plane, which fully defines the type-2
  antiferromagnetic structure. The dark and light lines represent
  the strong Cu-O-W-O-Cu $J_2$ exchange interactions, forming the two
  independent Cu$^{2+}$ sublattices with dark and light circles. The dark
  sublattice is perfectly AF ordered. On the light
  (crystallographically equivalent) sublattice there is a phase shift
  causing a domain wall (broken line). At the domain wall on the light
  sublattice, the composite 2D magnetic wave vector $k$ points along
  the $a$-axis (right) and at the domain boundary it rotates by
  90$^{\circ}$, to point along the $b$-axis ($k'$).
\label{fig:lattices}} 
\end{figure}

The magnetic structure implied by the $\left({}\frac{1}{2}, 0,
\frac{1}{2}\right){}$ peak at 0.68~\AA$^{-1}$ is the so-called type-2
structure, indexed by $k=\left({}\frac{1}{2}, 0, \frac{1}{2}\right){}$
(magnetic space group $P_s\bar{1}$). This structure may be
described as a series of antiferromagnetic planes stacked
ferromagnetically along the $b$ direction
(Fig.~\ref{fig:lattices}). As for the Ba analogue, the ordered moment
is very small, 
$\mu_{\textrm{o}} = 0.15\pm0.03\mu_{\textrm{B}}$, and its orientation
could not be determined reliably.

While Vasala \emph{et al.} did not observe any magnetic Bragg peaks in their
low-temperature neutron diffraction data, they did correctly predict
the magnetic structure to be of type-2 with the aid of \emph{ab-initio}
calculations combined with X-ray absorption-spectroscopy
data~\cite{vasala_2014_characterization}. 

\subsection{Muon spin resonance}
\begin{figure}
\includegraphics[width=8.6cm]{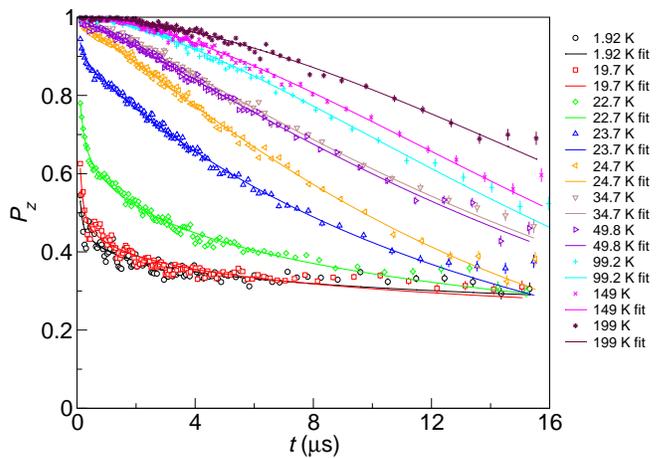}
\caption{MuSR on \SCWO{} in zero field. Fits are to a linear
  combination of stretched exponential and fast Lorentzian decay
  (Equation \ref{eqn:muSRFits}). \label{fig:SCWOmuSRDepolarisation}} 
\end{figure}

The previous $\mu$SR study revealed a sharp magnetic transition and
yielded the field strength at the muon site as a function of temperature below $T_N =
24$~K~\cite{vasala_2014_characterization}. This transition is also
observed in our data (Fig.~\ref{fig:SCWOmuSRDepolarisation}) with a
sharp drop in the $t=0$ polarization between 25 and 20~K. The present
data, collected at the ISIS facility (which is a high-intensity
pulsed source) is complementary to the data by Vasala~\emph{et al.}
taken at the Paul Scherrer Institute (PSI); the time resolution at MuSR at ISIS is lower, but the
muon relaxation can be followed over much longer time scales (up to
$16~\mu$s) due to the much higher number of muons implanted with each
pulse/time frame. The muon source at PSI is a continuous source and
therefore better suited for measurements requiring a high time
resolution. 

Between 20 and 200~K the muon relaxation can be fitted with a
(phenomenological) linear combination of Lorentzian and stretched
exponential decays; 
   
\begin{figure}
\includegraphics[width=8.6cm]{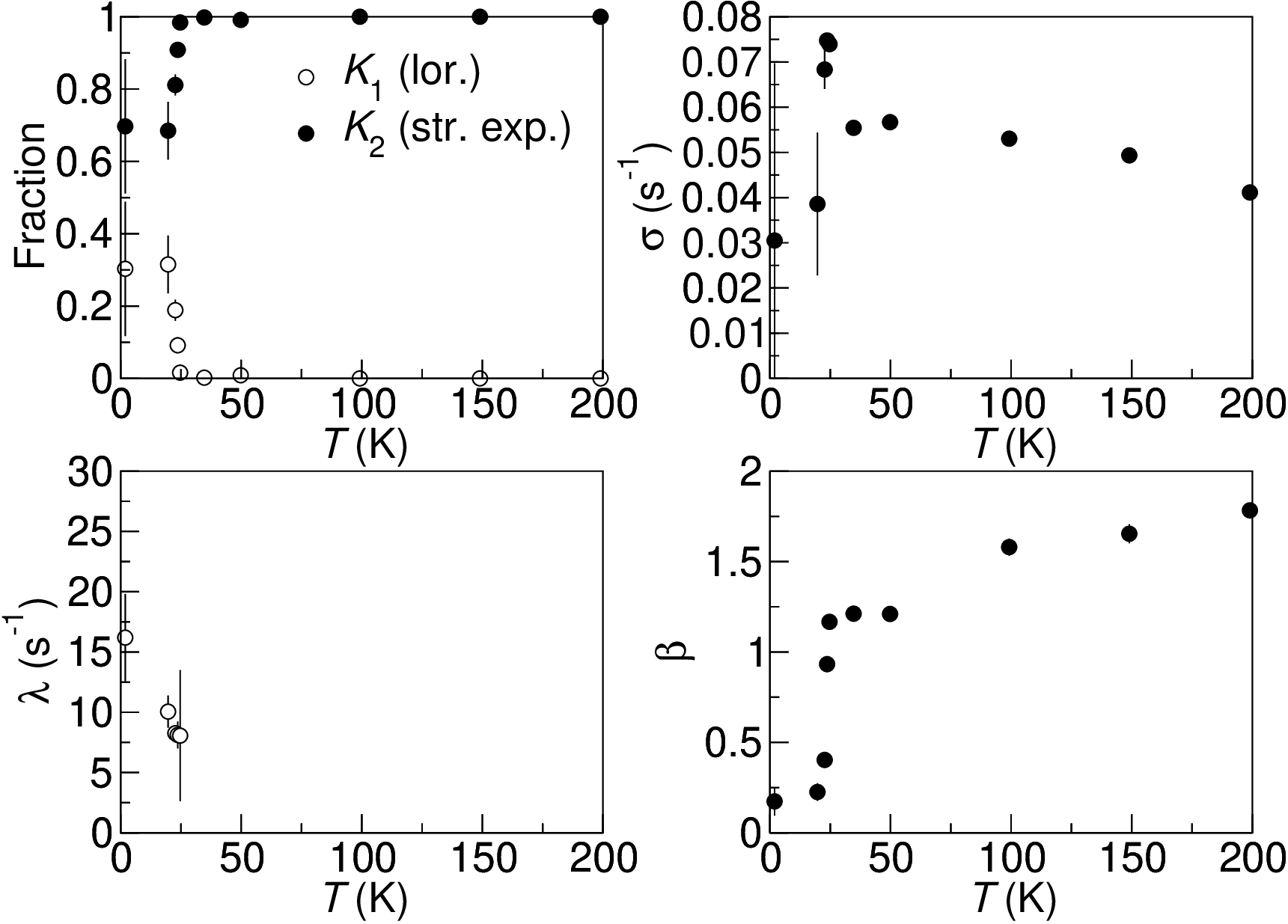}
\caption{Parameters to fits of muSR data for \SCWO{} (see
  Fig. \ref{fig:SCWOmuSRDepolarisation}).\label{fig:muSRParams}} 
\end{figure}

\begin{equation}
P_z = K_1\left({}e^{-\lambda{}t}\right){}+K_2\left({}e^{-\left({}\sigma{}t\right){}^{\beta{}}}\right){},
\label{eqn:muSRFits}
\end{equation}
with the additional constraint that
$K_1+K_2=1$. $\lambda{}$ and $\sigma{}$ are the Lorentzian and
stretched/compressed exponential relaxation factors respectively, and $\beta{}$
is the stretching exponent - in the present case, with $\beta>1$ the
relaxation is \emph{compressed} exponential, approaching $\beta
\approx 1.8$ at high temperatures; $\beta = 2$ would equate Gaussian
relaxation as expected for a paramagnet~\cite{aharenprb2009}.
The results of the fits at temperatures between $2$~K and $199$~K
are shown in Fig.~\ref{fig:muSRParams}. Down to 24.7~K,
right above the ordering transition, the Lorentzian component remains
close to zero and the relaxation is well modelled by the compressed
exponential. As the system is cooled from 200~K there is a steady
increase in $\sigma$ and a gradual reduction in $\beta$, providing
evidence of a gradual slow-down of the spin dynamics that starts at
temperatures far above $T_{\textrm{N}}$. 

Below $T_N=24$~K a significant contribution from fast Lorentzian
relaxation can be seen, and the stretching exponent $\beta$ drops
sharply to values below 1. Considering both the muon relaxation curves
of Fig.~\ref{fig:SCWOmuSRDepolarisation} and the fit data of
Fig.~\ref{fig:muSRParams} it is clear that while the two components in
Eqn.~\ref{eqn:muSRFits} accurately model two types of muon relaxation
present in the sample above $T_{\textrm{N}}$, this is no longer the case below
20~K. In the 3D-ordered state strongly damped oscillations are
observed, but in this case not with sufficient resolution to determine
the ordered moment as a function of temperature, as was done by Vasala
\emph{et al.}~\cite{vasala_2014_characterization}. The strong
damping of the oscillations points to a wide distribution of local
fields around the non-zero mean value. This is indicative of a fair
amount of magnetic disorder, which may be due to the presence of many
small magnetic domains, as we will see later. The relaxation to
1/3 of the initial polarisation shows that
one-third of the moments are aligned
parallel or antiparallel with the initial muon polarisation, as
expected for static moments in a powder sample with no preferential
crystallographic and therefore magnetic orientation.

\subsection{Inelastic neutron scattering}
\begin{figure}
\includegraphics[angle=270,width=8.6cm]{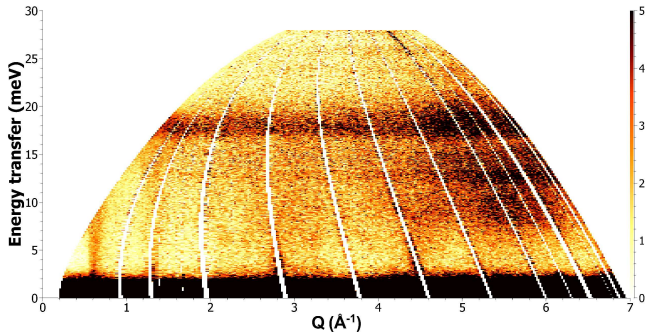}
\caption{Full inelastic neutron scattering spectrum of \SCWO{} at
  $5$~K.\label{fig:5Kfullcolourmap}} 
\end{figure}

\begin{figure*}
\includegraphics[width=17.2cm]{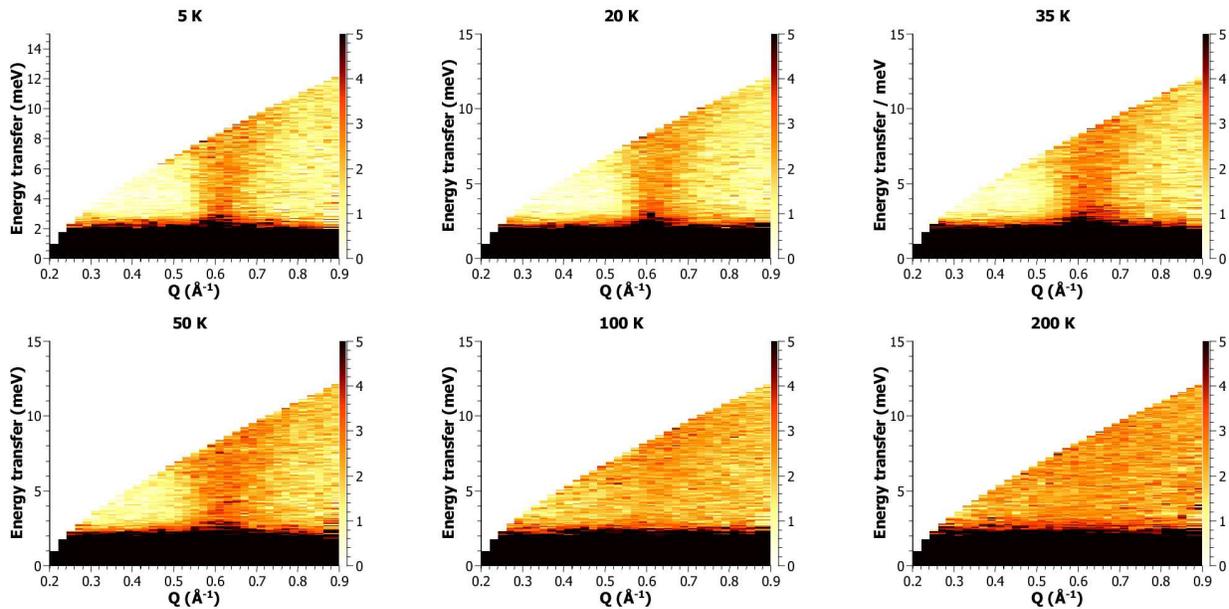}
\caption{Inelastic neutron scattering spectrum of \SCWO{} at a range
  of temperatures, showing strong scattering at
  $Q=0.63$\AA{}$^{-1}$ below
  $100$~K.\label{fig:colourmap}} 
\end{figure*}

Inelastic neutron scattering measurements were carried out to
characterise the dynamic magnetic correlations as a function of
temperature. 30~meV neutrons were used ($\lambda = 1.65~$\AA) and the
neutron spectrum taken at 5~K is shown in
Fig.~\ref{fig:5Kfullcolourmap}. The two bands below $1.5~$\AA$^{-1}$,
rising up to approximately 14~meV, are of magnetic origin -- no
structural reflections occur at these wave numbers. We focus in
particular on the part of the magnetic dispersion rising up from
approximately $Q=0.68$~\AA$^{-1}$, where the $\left({}\frac{1}{2}, 0,
\frac{1}{2}\right){}$ reflection is located at the elastic line. In
Fig.~\ref{fig:colourmap} the inelastic scattering in this area is
shown as a function of temperature. 

\begin{figure}
\includegraphics[width=8.6cm]{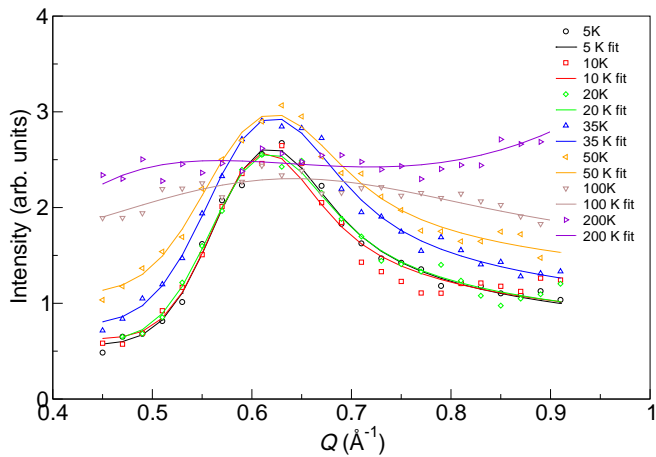}
\caption{Warren function (Equation \ref{eqn:Warren}) fit to inelastic
  peak below $100$~K, integrated between $4$~meV and
  $6$~meV.\label{fig:Warren}} 
\end{figure}

The energy window for neutrons at these short wave numbers is rather
narrow regardless of the energy of the incident neutrons, but with
30~meV neutrons a significant section of the magnetic spin-wave
dispersion and dynamic magnetic correlations can still be
observed. The zero-energy-transfer origin of the dispersion, and
center of mass, appears to be at about 0.62~\AA$^{-1}$, somewhat below
the elastic magnetic reflection at 0.68~\AA$^{-1}$.
Figure~\ref{fig:Warren} shows the neutron scattering intensity
integrated over energies from 4 to 6~meV and corrected for the
magnetic form factor, showing the same mismatch. The reason for the
mismatch is that the dispersion observed here does \emph{not}
correspond to the 3D spin-waves in the \emph{ordered} phase. The
energy scale of the 3D ordering is up to $\sim 2$~meV and the signal
up to that energy is swamped by the elastic line and hence is
invisible here. The dispersion observed here corresponds to 2D dynamic
antiferromagnetic correlations - the energy scale of 14~meV ($\approx
162$~K) is of the 
order of the in-plane magnetic exchange interactions. It is then not
surprising that the dispersion curves remain practically unchanged on
heating above the ordering transition, with strong correlations
visible even at 50~K and still just about visible at 100~K. 

\begin{figure}
\includegraphics[width=8.6cm]{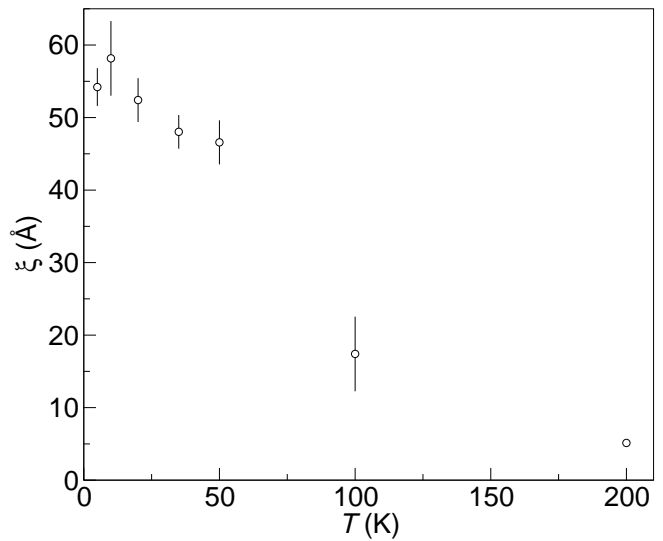}
\caption{Change in correlation length in Warren function with
  temperature.\label{fig:WarrenXi}} 
\end{figure}

The energy-integrated scattering shown in
Fig.~\ref{fig:Warren} exhibits a peak shape that is
characteristic of 2D correlations in a powder sample, as described by
Warren~\cite{warren_1941_x-ray,wills_1999_two-dimensional};
\begin{equation}
I=\frac{C\left({}1-2\left({}\frac{\lambda{}Q}{4\pi{}}\right){}^2+2\left({}\frac{\lambda{}Q}{4\pi{}}\right){}^4\right){}\sqrt{\frac{\xi{}}{\lambda{}\sqrt{\pi{}}}}
  \times \int_{0}^{20}\!F(a)\,\mathrm{d}x}{\left({}\frac{\lambda{}Q}{4\pi{}}\right){}^{\frac{3}{2}}}+K
\label{eqn:Warren}
\end{equation}
where
\begin{equation}
F(a)=e^{-\left({}x^2-a\right){}^2}
\label{eqn:WarrenF}
\end{equation}
and
\begin{equation}
a=\frac{\xi{}\sqrt{\pi{}}\left({}Q-Q_0\right){}}{2\pi{}}
\label{eqn:WarrenF2}
\end{equation}
and $\lambda = 1.65~$\AA{} is the neutron wavelength. The wave vector obtained from
the fit $Q_0~=~0.592\pm0.002$~\AA$^{-1}$  corresponds to the in-plane
$\left({}\frac{1}{2},0,0\right){}$ antiferromagnetic wave
vector. Strickly speaking, the Warren function (Eq.~\ref{eqn:Warren})
describes only the \emph{elastic} scattering from short-ranged 2D
correlations. In the present case we integrate over a small range of
(relatively-low) inelastic energies as is in practice inevitable in
diffuse neutron scattering. Here it is justified by the energy
independence of the neutron-scattering cross section within and, where
visible, outside the interval of energies included in the integral.
Hence this data confirms that the magnetic correlations above
$T_{\textrm{N}}$ are 2D (in the $a-b$ planes), and persist, if weakly,
up to 100~K. The fits yield a lower limit to the correlation length
$\xi$ as a function of temperature, shown in Fig.~\ref{fig:WarrenXi}.  

\section{Discussion}

The presence of 2D dynamic antiferromagnetic correlations
above $T_{\textrm{N}}$ and up to $\sim 100$~K, as observed in the neutron spectra
of Figs.~\ref{fig:colourmap} and \ref{fig:Warren}, explains the absence
of an anomaly at $T_{\textrm{N}}$ in the magnetic susceptibility. We can conclude
that below the broad maximum in the magnetic susceptibility data
(Fig.~\ref{fig:SCWOSusc}) at 100~K antiferromagnetic spin correlations
start to grow as the temperature is lowered, while not far above 100~K
Curie-Weiss paramagnetism sets in. The $\mu$SR data show that there
is a gradual slowing down of spin fluctuations with increasing 2D
spatial correlations, but that the spins remain dynamic until cooled
below $T_{\textrm{N}}$. 

These observations are also consistent with the very small peak that
can just be discerned in the heat capacity data between 20 and 26~K,
peaking at $\sim 24$~K. The (2D) magnetic wave vector in the 2D layers
is $\left( \frac{1}{2}, 0 \right)$ and this leads to three distinct
ways an ordered 2D layer can be stacked on the preceding one along the
$c$-axis: shifted by 1/2 a phase along the magnetic $k$-vector, and
rotated by 90$^{\circ}$ around the $c$-axis. The entropy associated
with the 2D- to 3D-ordering  transition can then be calculated as
follows: the total number of distinct configurations of $l$ 2D ordered
layers  $\Omega(l)=3^{l-1}$. Hence the entropy of the 2D- to
3D-ordering transition as a function of $l$ is given by
\begin{equation}
S_{\textrm{2D-3D}}(l) = \ln[\Omega(l)] = \ln(3^{l-1}) = (l-1) \ln(3) ~~\propto n^{\frac{1}{3}}\ln(3),
\end{equation}
where $n$ is the total number of magnetic $S=1/2$ spins in a sample
grain or in the coherent volume. Note that we see from the approximate
equation at the right-most side that this entropy is subextensive.  If
we assume that the coherent volume below the transition contains an
equal number of spins along each crystallographic axis then the total
magnetic entropy of that volume can be written as
\begin{equation}
S_{\textrm{magnetic}}(l)= l^3\ln(2)
\end{equation}
and the ratio
\begin{equation}
S_{\textrm{2D-3D}}(l) / S_{\textrm{magnetic}}(l) = \frac{l-1}{l^3}\frac{\ln(3)}{\ln(2)}
\end{equation}
The experimentally-obtained ratio of 0.003 (0.3\%) (an upper limit)
yields $l=15$, which translates to an upper limit for the correlation
length of 63~\AA~ along the $c$-axis and 81~\AA~ in the $a-b$
plane. This is reassuringly close to the correlation length in the
$a-b$ plane just above the  transition (55~\AA), as obtained from the
Warren fits. It is also consistent with the fast muon relaxation to
$P_z = 1/3$ below $T_{\textrm{N}}$, which points to small coherent
antiferromagnetic domains in the low-temperature 3D-ordered
phase. Given the degeneracy intrinsic to the type-2 magnetic ordering
discussed earlier, these are likely to be domains, with the in-plane
component of the magnetic $k$ vector variably aligned along the
crystallographically-equivalent $a$- and $b$-axes.

The available data are fully consistent and readily explained by the
presence of strong 2D correlations persisting above $T_{\textrm{N}}$, in what
could be described as a ``2D thermal spin liquid'' phase at intermediate
temperatures. While this is in line
with earlier proposals~\cite{vasala_2014_characterization} it is not
clear how this phase arises if the computational exchange
constants~\cite{vasala_2014_characterization} are correct -- in
particular if $J_4$ is indeed as strong as $-4.21$~meV (49~K), 56\% of
the dominant $J_2$ exchange. This would not be a quasi-2D system and could be
expected to show 3D \Neel{} ordering around at least 50~K. Note that the two
interpenetrating 3D sublattices arising from the $J_2$ and $J_4$ bonds
only (Fig.~\ref{fig:lattices}) are not by themselves
frustrated. When they order fully antiferromagnetically, each with a
wave vector $\left( \frac{1}{2}, \frac{1}{2}, \frac{1}{2} \right)$, 
there is a two-fold degeneracy arising from the relative phase of the
two sublattices. Both degenerate states are type-2 
antiferromagnetic, with the in-plane antiferromagnetic phase vector
aligned along the $a$ or the $b$ axis. This
degeneracy is not lifted by inclusion of the much weaker $J_1$ and
$J_3$ because for each $J_1$ edge with antiferromagnetically aligned
spins at the vertices there is a $J_1$ edge with ferromagnetically
aligned spins. 

Hence, the observed type-2 antiferromagnetism is consistent with the
predicted exchange constants, but the dynamic correlations in the 2D
thermal spin liquid state above $T_{\textrm{N}}$ are not.  In the
archetypical quasi-2D antiferromagnet \ce{La2CuO4} the ratio between
the interplane exchange constant and the main in-plane exchange
interaction $J_{\perp}/J_1 = 10^{-5}$. This is hugely different from
the ratio between the predicted $J_4$ and $J_2$ for \SCWO, considering
that the ratio of the \Neel temperature $T_{\textrm{N}}$ over $J_1$ in
\ce{La2CuO4}; 300~K / 1345~K~$= 0.22$, is only slightly less than
$T_{\textrm{N}}$ over $J_2$ in \ce{Sr2CuWO6}. Furthermore, the
short-ranged correlations evident above $T_{\textrm{N}}$ in
\ce{Sr2CuWO6} are 2D and strongly reminiscent of the magnetism in
\ce{La2CuO4} above 300~K.  Hence, it is very unlikely that the
theoretically-predicted value for $J_4 = -4.21$~meV is correct. 

One possible explanation for the computational overestimate lies in
the role of the W 5$d^1$ orbital. The W cations are part of the $J_2$
and $J_4$ interaction pathways and XAS by Vasala~\emph{et
  al.}~\cite{vasala_2014_characterization} indicates that there is some
overlap between the Cu~3$d$~--~O~2$p$ bands and the W~5$d$~--~O~2$p$ bands. It may
then be of relevance that the magnetism of W~5$d^1$ electrons in an
octahedral crystal field is described by a Kugel-Khomski Hamiltonian
in the limit of strong spin-orbit coupling, as described in detail by
Gang-Chen~\emph{et al.}\cite{chen_2010_exotic}. This gives rise to a magnetic
Hamiltonian in which the exchange pathway itself depends on the
orientation of the magnetic (spin-orbital) moments. Combined with the
solid orbital ordering of the magnetic Cu$^{2+}$ $d_{x^2-y^2}$ orbital
in the $a-b$ plane, perhaps the strongly spin-orbit-coupled magnetism of
the W cation causes a much greater difference between $J_2$ and $J_4$
interactions. This is something that might be worth exploring in
further detail. Further neutron spectroscopy experiments, from which
the exchange constants could be determined via spin-wave modelling,
would also be beneficial but probably only if done using large single
crystals which are presently not available.

With regards to the muon relaxation in the thermal spin liquid regime, 
it is worth noting that compressed exponential relaxation has been
observed in other spin liquid phases, for example in
\ce{SrCr8Ga4O19} (SCGO)~\cite{uemura_1994_spin}, herbertsmithite
\ce{(ZnCu3(OH)6Cl2)}~\cite{mendels_2007_quantum} and the double
perovskite \ce{Ba2YMoO6}~\cite{devries_2013_low-temperature}. In SCGO
the compressed exponential approaches Gaussian ($\beta =2$) as the
temperature is lowered and it is thought 
to arise from a dynamical and heterogeneous ground state with a
mixture of unpaired spins and spins paired into spin-singlet
dimers~\cite{uemura_1994_spin}, so that any particular muon environment fluctuates between
magnetic and non-magnetic at a time scale shorter than the muon
lifetime. This inevitably gives rise to a lifetime faster than that
observed for the other examples and in \ce{Sr2CuWO6} there is no
evidence of spin-singlet dimers, as they would be manifest in neutron
spectroscopy data as near-neighbour antiferromagnetic correlations
only. This also rules out any fundamental similarity with the spin
liquid states in herbertsmithite and \ce{Ba2YMoO6} as both states are
thought to be dominated by near-neighbour
spin-singlet~\cite{devries_2009_scale-free} and spin-orbital singlet
dimers~\cite{devries_2013_low-temperature}, respectively. In
\ce{Ba2YMoO6} the compressed exponential muon relaxation is the result
of two distinct muon environments: those adjacent to
unpaired (dangling) fluctuating spins and those sitting elsewhere in
the material, surrounded by non-magnetic spin-orbital singlet
pairs. Hence there is a rather large variety of scenarios that give
rise to compressed exponential relaxation and they have to be
considered case by case. No further parallels can be drawn between the
case of \ce{Sr2CuWO6} and the others. 

\section{Conclusions}
Below 24~K the quasi-2D square lattice antiferromagnet \SCWO{} freezes
into a 3D-ordered N\'{e}el state with k-vector \AFReflection{},
pointing to type-2 antiferromagnetic order (magnetic space group
$P_{\textrm{s}}\overline{1}$) as predicted previously  based on X-ray
absorption spectroscopy combined with computational
studies~\cite{vasala_2014_characterization}. As often for quasi-2D
antiferromagnets with $S=1/2$, neutron spectroscopy shows that dynamic
2D correlations are detectable far above $T_{\textrm{N}}$; in this case up to
$\sim 100$~K, a temperature that is comparable to the energy scale of
the main exchange interaction estimated at $J_2 \sim - 90$~K and
coinciding with the maximum in the magnetic
susceptibility. Furthermore, $\mu$SR shows a pronounced slowing down
taking place well before $T_{\textrm{N}}$ is reached on cooling.  

The pronounced 2D character of the magnetism is however surprising in
the light of the exchange constants one might expect considering the
structure of \SCWO. Perhaps the spin-orbit coupling in the
5$d^1$ band of the W cations on the dominant exchange interaction
pathway, not taken into account in the reported
calculations~\cite{vasala_2014_characterization}, can explain the
large difference between in-plane and out-of-plane exchange
interactions implied by the above observations.

\begin{acknowledgments}
The authors thank EPSRC for Ph.D. student funding (OB), and the STFC and ILL for beam time.
\end{acknowledgments}

%\bibliography{ZotOutput}

\begin{thebibliography}{27}
\bibitem [{pos()}]{positrons}%
  \BibitemOpen
  \href@noop {} {\ }\bibinfo {note} {Positrons are emitted as a muon decay
  product. The direction of their emission is preferentially along the
  direction the muon spin pointed in at the time that muon
  decayed.}\BibitemShut {Stop}%
\end{thebibliography}

\end{document}